\documentstyle[12pt]{article}

\advance\textwidth1in
\advance\oddsidemargin-.5in
\evensidemargin\oddsidemargin
\advance\topmargin-1in
\advance\textheight1.5in

\newtheorem{theorem}{Theorem}[section]
\newtheorem{claim}[theorem]{Claim}
\newtheorem{definition}[theorem]{Definition}
\newtheorem{corollary}[theorem]{Corollary}
\newtheorem{lemma}[theorem]{Lemma}

\def\qed{\nopagebreak[4]\par\rightline{\fbox{\phantom{i}}}\par\bigskip\par}

\def\C{\hbox{${\bf C}\!\!\!\!\hbox{I}\,$}}
\def\P{\hbox{$\hbox{I}\!\bf P$}}
\def\N{\hbox{$\hbox{I}\!\bf N$}}

\def\supp{\hbox{supp }}
\def\t{\underline{\bf t}}

\def\Zc{\cal}
\def\calg{{\it g}}
\def\dC{{\Zc C}}
\def\M{{\Zc M}}

\def\ltag#1$${\leqno\hbox{$(#1)$}$$}
\def\ttwo{t_2}
\def\Ttwo{T_2}
\def\CM{{\Zc CM}}
\def\Mxi{\xi}
\def\Meta{\eta}
\def\BISxi{{\zeta}}
\def\BISeta{{\mu}}
\def\cmps{Calogero-Moser particle system}
\def\cmpss{{\cmps}s}
\def\stat{\psi_W}%
\def\iseries#1{\,[\![#1]\!]}

\date{hep-th/9412124 \\\small(to appear in {\it Communications in
Mathematical Physics})}

\title{Bispectral KP Solutions and Linearization of Calogero-Moser
Particle Systems}

\author{Alex Kasman\thanks{Research supported by NSA Grant
MDA904-92-H-3032} \\ {\it Department of Mathematics} \\ {\it Boston
University}}

\begin{document}

\maketitle

\begin{abstract} Rational and soliton solutions of the KP hierarchy in
the subgrassmannian $Gr_1$ are studied within the context of finite
dimensional dual grassmannians.  In the rational case, properties of
the tau function, $\tau$, which are equivalent to bispectrality of the
associated wave function, $\psi$, are identified.  In particular, it
is shown that there exists a bound on the degree of all time variables
in $\tau$ if and only if $\psi$ is a rank one bispectral wave
function.  The action of the bispectral involution, $\beta$, in the
generic rational case is determined explicitly in terms of dual
grassmannian parameters.  Using the correspondence between rational
solutions and particle systems, it is demonstrated that $\beta$ is a
linearizing map of the Calogero-Moser particle system and is
essentially the map $\sigma$ introduced by Airault, McKean and Moser
in 1977 \cite{AMcM}.  \end{abstract}

\section{Introduction}

Among the surprises in the history of rational solutions of the KP
hierarchy (and the PDE's which make it up) are the existence of
rational initial conditions to a non-linear evolution equation which
remain rational for all time \cite{AM,AMcM}, that these solutions are
related to completely integrable systems of particles
\cite{AMcM,Kr1,Kr}, and that a large class of wave functions which
have been found to have the bispectral property turn out to be
associated with potentials that are rational KP solutions
\cite{DG,W,Z}.  Within the grassmannian which is used to study the KP
hierarchy, the rational solutions, along with the $N$-soliton
solutions, reside in the subgrassmannian $Gr_1$ \cite{SW}.  This paper
develops a general framework of finite dimensional grassmannians for
studying the KP solutions in $Gr_1$ and then applies this to the
bispectral rational solutions.  New results include information about
the geometry of KP orbits in $Gr_1$ and identification of properties
equivalent to bispectrality.  In addition, an explicit description of
the bispectral involution in terms of dual grassmannian coordinates
leads to the conclusion that it is, in fact, essentially the
linearizing map $\sigma$ \cite{AMcM}.

\subsubsection{Outline}

A brief review of previous results which will be necessary to
understand the results of this paper follows in
Section~\ref{sec:history}.  For details, please refer to \cite{KrN,Pr}
for a review of the theory of spectral curves, \cite{Mu} for the
algebraic theory of the KP hierarchy and to \cite{SW} for the analytic
theory of the grassmannian which will be used in this paper.

Section \ref{sec:dual} develops the technique of dual grassmannians
for studying the $Gr_1$ solutions of the KP hierarchy.  This
construction takes advantage of the fact that the orbits of subspaces
in $Gr_1$ are infinite dimensional subspaces in which each point has a
finite dimensional complement.  Consequently, the dual grassmannians
are finite dimensional and simpler to work with.  The decomposition of
the dual grassmannians into the disjoint union of generalized
jacobians and the representation of their Pl\"ucker coordinates as a
$\tau$ function are discussed.

In Section \ref{sec:bisp}, the technique of the dual grassmannian is
utilized to study bispectrality of the wave function.  Using results
of Wilson \cite{W} which completely identified the bispectral wave
functions in $Gr_1$, a set of conditions on $W$ or $\tau_W$ which are
equivalent to bispectrality of $\psi_W$ are listed
(Theorem~\ref{thm:bisp}).  In particular, it is shown that for $W\in
Gr_1$, the wave function $\psi_W$ is bispectral if and only if
$\tau_W$ is of {\it bounded degree\/} in all time variables.

The bispectral involution introduced by Wilson \cite{W} is studied in
Section~\ref{sec:beta}.  It is noted that the composition of the
bispectral involution with the KP flows results in a non-isospectral
flow under which the singular points of the spectral curve move as a
\cmps.  The action of the involution is determined explicitly
(Theorem~\ref{coords}) in terms of dual grassmannian parameters in the
generic case that the spectral curve of the image has only simple
cusps.  Using the correspondence between rational solutions and
particle systems, the bispectral involution is determined as an
involution on \cmpss\ (Theorem~\ref{image}), and it is shown
(Theorem~\ref{action-angle}) that it acts as a linearizing map.  The
bispectral involution acting on particle systems, after simple
rescaling, is seen (Theorem~\ref{sigma}) to restrict to the
linearizing map $\sigma$ introduced in \cite{AMcM} and thus explains
the surprising involutive nature of $\sigma$ in terms of
bispectrality.  The paper concludes with two examples.  The first
example demonstrates how the bispectral involution can be used to
determine the $\tau$-function of a rational KP solution.  Finally, the
second example demonstrates Theorem~\ref{coords} by determining the
condition space of the image under the bispectral involution of a
particular rational solution of the form given by Wilson \cite{W} in
Example 10.3.

\subsection{Review}\label{sec:history}%

\subsubsection{Rational Solutions to the KP Equation}

For the purposes of this paper, the term ``rational solution of the KP
Equation'' will mean a function $u(x,y,t)$ which is rational in the
variable $x$, satisfies the KP equation $$ \frac34
u_{yy}=\left(u_t-\frac14(6uu_x+u_{xxx})\right)_x $$
 and has the property that $\lim_{|x|\to\infty}u=0$.  Although there
do exist rational solutions which do not have the latter property
\cite{Gr,V}, these will only be referred to explicitly as
``non-vanishing rational solutions''.

Along with the ideas of the algebro-geometric construction \cite{KrN},
the correspondence between poles of rational solutions and the
Calogero-Moser system \cite{AMcM} is utilized by Krichever
\cite{Kr1,Kr} to find rational solutions to the KP equation.  The
motion of the poles is shown to be equivalent to the motion of
Calogero-Moser particles where the $y$-flow corresponds to the second
Hamiltonian and the $t$-flow corresponds to the third.  Furthermore,
this paper completely identified the rational solutions of the KP
equation.

The following results from \cite{Kr} will be important in the present
paper.

\begin{theorem}[Krichever]\label{Kr1} A function $u(x,y,t)$ which is
rational in $x$ and decreases for $|x|\to\infty$ is a solution of the
KP equation if and only if $$
u(x,y,t)=\sum_{j=1}^N\frac{-2}{(x-x_j(y,t))^2} $$ and there exists a
function $$
\psi(x,y,t,z)=\left(1+\sum_{j=1}^N
\frac{p_j(y,t,z)}{(x-x_j(y,t))}\right)e^{xz+yz^2+tz^3}
\ltag B $$ such that $$
L_1\psi=\frac{{\partial}}{{\partial}y}\psi,\qquad
L_2\psi=\frac{{\partial}}{{\partial}t}\psi $$ $$
L_1=\frac{{\partial}^2}{{\partial}x^2}+u(x,y,t),\qquad
L_2=\frac{{\partial}^3}{{\partial}x^3}
+\frac{3}{2}u\frac{{\partial}}{{\partial}x}+w(x,y,t)
$$ $$
w(x,y,t)=\sum_{j=1}^N\left[3(x-x_j)^{-3}
+\frac{3}{2}(x-x_j)^{-2}\frac{{\partial}}{{\partial}y}x_j\right].  $$
\end{theorem}

In general, a wave function $\psi$ corresponding to the solution $u$
is any function of the form
$$\left(1+\alpha_1(x,y,t)z^{-1}
+\alpha_2(x,y,t)z^{-2}+\cdots\right)e^{xz+yz^2+tz^3}
$$ which satisfies $L_1\psi=\frac{{\partial}}{{\partial}y}\psi$ and
$L_2\psi=\frac{{\partial}}{{\partial}t}\psi$ for $L_1$, $L_2$ and $u$
as in Theorem~\ref{Kr1}.  Multiplication by any series of the form
$1+c_1z^{-1}+c_2z^{-2}+\cdots$ (with $c_i\in\C$) will take one wave
function for $u$ to another wave function for the same $u$.  Since a
series of this form can be seen to alter $\psi$ but not the associated
solution $u$, they will be referred to as {\it gauge
transformations\/} of the wave function.  As will be shown below by
Corollary~\ref{limx}, there is a unique wave function of the form
$(B)$ corresponding to each rational solution.  This form of the wave
function is the {\it bispectral gauge\/} and can be identified by the
fact that $\lim_{x\to\infty}exp(-xz-yz^2-tz^3)\psi=1$.

A method for generating almost every rational solution of the KP
equation is developed (also in \cite{Kr}) utilizing {\it another\/}
gauge of the wave function, which is identified by the fact that
$exp(-xz-yz^2-tz^3)\psi|_{x=0}=1$.  In this gauge, denoted $(Kr)$,
there exist $N$ distinct numbers $\lambda_i$, such that
$\frac{{\partial}}{{\partial}z}\psi|_{z=\lambda_i}=0$.  Consequently,
it is clear that the spectral curve in this case is a rational curve
with singularities only in the form of simple cusps at the points
$z=\lambda_i$.  In the standard algebro-geometric construction of
solutions to integrable equations with a {\it non\/}-singular spectral
curve, the wave function $\psi$ is chosen to be an {\it Akhiezer
function\/}.  That is, it is chosen so as to be the unique wave
function which is holomorphic off of a given non-special divisor and
specified point.  The wave function in the gauge $(Kr)$ is the limit
of the Akhiezer functions in the cuspidal case.  It is for its
relevance to the algebro-geometric construction that this gauge is
introduced here, although it is the bispectral gauge that will be
utilized in the proofs to follow.

The wave function of a rational KP solution can be written in the form
$$
\psi(x,y,t,z)=\left(1+\frac{p(x,y,t,z)}{q(z)}\right)e^{xz+yz^2+tz^3},
$$ where $q$ is a monic polynomial of degree $N$ and $p$ is a
polynomial of degree at most $N-1$ in $z$.  There is a unique such
function in the gauge $(Kr)$ for any choice of the $\lambda_i$ and the
coefficients of $q$ and thus these parameters determine a KP solution
$u$.  After specifying $\lambda_i$ and $q$, the problem of finding
$\psi$ reduces to a problem of linear algebra.  The matrix, $\Theta$,
which arises in this problem leads to a solution to the KP equation as
shown in this theorem.

\begin{theorem}[Krichever]{\label{Kr2} For almost all solutions of the
KP equation, depending rationally on $x$ and decreasing for
$|x|\to\infty$, we have the formula $$
u(x,y,t)=2\frac{{\partial}^2}{{\partial}x^2}\log\det\Theta $$ where
the matrix elements $\Theta_{ij}$ are given by\footnote{Whereas the
matrix $\Theta$ in \cite{Kr} was indexed from 0 to $N-1$, I have
altered the notation to be compatible with the matrix $\M$ below.  In
addition, the substitution $t\to-t$ is understood when comparing the
present paper to \cite{Kr}.}  $$
\Theta_{ij}=\frac{{\partial}}{{\partial}z}
\frac{z^{j-1}e^{xz+yz^2+tz^3}}{q(z)}
\bigg|_{z=\lambda_i}e^{-(x\lambda_i+y\lambda_i^2+t\lambda_i^3)}\qquad1\leq
i,j \leq N $$} \end{theorem} \noindent Since the elements of $\Theta$
are all linear in $x$, $y$ and $t$, letting $\vartheta=\det\Theta$ we
have \begin{corollary}\label{polyxyt}{Almost every rational solution
can be expressed as $2\frac{{\partial}^2}{{\partial}x^2}
\log\vartheta$ where $\vartheta$ is a polynomial in $x$, $y$ and $t$.}
\end{corollary}

\noindent{\bf Note:} Those solutions which can be determined
explicitly from Theorem~\ref{Kr2} are those for which all of the cusps
are {\it simple\/}.  Rational solutions whose spectral curves have
higher cusps can be determined as limits of these solutions.

\subsubsection{The KP Hierarchy and Sato's Grassmannian}

The KP hierarchy is defined in terms of pseudo-differential operators
in the variable $x$.  Given a pseudo-differential operator
$L=\sum_{i=-\infty}^{1} a_i(x)\partial^i$ where
$\partial=\frac{\partial}{\partial x}$ and $a_1(x)=1$, one may define
the flows of the KP hierarchy as the compatibility of the conditions
$$ L\psi=z\psi \qquad\frac{\partial}{\partial t_i}\psi=(L^i)_+\psi
\ltag KP $$ with $(K)_+$ denoting the projection of $K$ onto ordinary
differential operators.  Here and throughout the paper, I will use the
notation $x=t_1$, $y=t_2$ and $t=t_3$.

The geometry of an infinite dimensional grassmannian, $Gr$, is
utilized to study the dynamical system of pseudo-differential
operators given above.  This is a grassmannian of the Hilbert space
$H$ which is spanned by all integer powers of the variable $z$.  Thus,
a point $W\in Gr$ is a subspace of $H$ (satisfying certain conditions
which will not be emphasized here).  As will be seen below, the action
of a certain multiplicative group on $Gr$ results in the KP flow of
the associated operators.

The association between a point $W$ and its corresponding
pseudo-differential operator, $L_W$, is described most easily in terms
of certain intermediate objects.  Associated to a point $W\in Gr$ is
the stationary wave function $\stat(x,z)$ which is the unique function
of the form $$
\stat(x,z)=\left(1+\alpha_1(x)z^{-1}+\alpha_2(x)z^{-2}\cdots\right)e^{xz}
$$ that is contained in $W$ for each fixed value $x$ in its domain.
This function can also be written as $K_We^{xz}$ for some monic, zero
order, pseudo-differential operator $K_W$.  Finally, we have the
associated pseudo-differential operator $L_W=K_W\partial (K_W)^{-1}$.
Also frequently associated to $W$ are the time dependent version
$\stat$ given by the equations $(KP)$, the tau function $\tau_W$ (a
``bosonic'' representation of the Pl\"ucker coordinates of $W$), and
the function $u_W(t_1,t_2,\ldots)$ which solves the original KP
equation.  Depending on the context, any of these associated objects
could be considered a ``solution'' of the KP hierarchy.  Again, for
the purposes of this paper, a rational solution of the KP hierarchy is
a solution such that the function $u_W$ is a rational solution of the
KP equation in the sense discussed earlier.  Although there is a one
to one correspondence between points $W\in Gr$ and the objects
$\tau_W$, $K_W$ or $\psi_W$, the map to $L_W$ and $u_W$ is many to
one.

The association of a $\tau$-function to a point of the grassmannian
comes through the representation of $Gr$ on $\P(\C\iseries{t_i})$.  It
is well known that the set of Schur polynomials form a basis over $\C$
of $\C\iseries{t_i}$.  In the representation of $Gr$, a Schur
polynomial is associated in the standard way to each of the points of
$Gr$ where all but one of the Pl\"ucker coordinates are zero.  Let $W$
be an arbitrary point of $Gr$, and let $\tau_W$ be the infinite series
written as a sum of the Schur functions with the corresponding
Pl\"ucker coordinates as the coefficients.  That is, if
$\pi_{\nu}\in\C$ are the Pl\"ucker coordinates of a point $W\in Gr$,
then $$ \tau_W=\sum_{\nu}\pi_{\nu} S_{\nu}(t_i) $$ where $S_{\nu}$ is
the Schur polynomial associated to $\pi_{\nu}$.  The function
$u(t_i)=2\frac{{\partial}^2}{{\partial}x^2}\log\tau_W$ is then a
solution to the KP equation.  Since it is exactly those coefficients
which are valid Pl\"ucker coordinates that yield solutions, the KP
hierarchy can be viewed as the Pl\"ucker relations of the
grassmannian.

Two groups of interest to this paper act on $Gr$ via multiplication.
The group $\Gamma_+$ consists of all real-analytic functions
$f:S^1\to\C^{\times}$ which extend to holomorphic functions
$f:D_0\to\C^{\times}$ in the disc $D_0=\{z\in\C:|z|\leq1\}$ satisfying
$f(0)=1$.  The other group, $\Gamma_-$ consists of functions $f$ which
extend to non-vanishing holomorphic functions in
$D_{\infty}=\{z\in\C\cup\infty:|z|\geq1\}$ satisfying $f(\infty)=1$.
If $g=e^{-cz^i}\in\Gamma_+$ and $W\in Gr$, then
$$\tau_{gW}(t_1,t_2,\ldots,t_i,\ldots)=\tau_W(t_1,t_2,\ldots,t_i+c,\ldots).$$
Thus, the KP flows are given by the action of the group $\Gamma_+$.
However, the group $\Gamma_-$, since it contains only elements of the
form $1+\alpha_1z^{-1}+\alpha_2z^{-2}\ldots$, is again a group of
gauge transformations which alter the point, $W$, but not the
associated $u_W$ or $L_W$.

The solution $u_W$ corresponding to a point which only has a {\it
finite\/} number of non-zero Pl\"ucker coordinates is going to be a
rational solution since $\tau_W$ (now a {\it finite\/} sum of Schur
polynomials) will be a polynomial.  The subset of $Gr$ for which this
is true is called $Gr_0$, and it corresponds to rational solutions
whose spectral curve is a rational curve with only one cusp, at the
point $z=0$.  It therefore generates some of Krichever's rational
solutions, and all of the rational KdV solutions.  This is contained
in a larger subset, $Gr_1$, which is characterized by the fact that if
$W\in Gr_1$, then there are polynomials $p(z)$ and $q(z)$ such that
$pH_+\subset W\subset q^{-1}H_+$.  (Here, $H_+$ denotes the subspace
of $H$ spanned by non-negative powers of $z$.) The points of $Gr_1$
correspond to rank one solutions with rational spectral curves and the
converse is true if the data are suitably normalized
\cite{SW}~(Proposition~7.1).  Thus, all of Krichever's rational
solutions can be derived from this sub-grassmannian.  However, not
every solution from $Gr_1$ is rational since it also includes the
nodal rational curves which are associated with
solitons\footnote{Although the rational solutions are technically
soliton solutions as well, within this paper I will refer only to
non-rational solutions as solitons.}.  Only solutions in $Gr_1$ will
be considered for the remainder of the paper.

It was shown in \cite{MuJac} that the orbit of a pseudo-differential
operator $L$ which is a solution of the KP hierarchy and can be
determined by an isospectral flow of line bundles is isomorphic to the
jacobian variety of the spectral curve.  Thus, for a point $W$ which
is determined by a line bundle over a curve, the orbit under
$\Gamma_+$ modulo the action of $\Gamma_-$ is isomorphic to the
jacobian of the curve.  In general, there is no reason to expect that
the orbit in $Gr$ will be a jacobian prior to taking the quotient by
$\Gamma_-$.  The next section will study the $\Gamma_+$ orbits of
points in $Gr_1$ and their relationship to the geometry of the
grassmannian.

\section{Dual Grassmannians
and Rational KP Solutions}\label{sec:dual}%

\subsection{Differential Conditions and
$Gr_1$}

Every subspace $W\in Gr_1$ can be derived from a line bundle over a
singular rational curve \cite{SW}.  As will be explained below, it can
also be written as the closure in $H$ of the set of polynomials in $z$
satisfying a finite number of differential conditions at a finite
number of points divided by a polynomial $q$ \cite{W}.  By forming the
grassmannian of the linear space of such conditions, one is able to
construct finite dimensional grassmannians dual to those in $Gr_1$.

\subsubsection{Definitions}

Let $d(l,\lambda)$ denote the linear functional on the space $\C[z]$
which takes $f(z)$ to $f^{(l)}(\lambda)$, the $l^{th}$ derivative
evaluated at $\lambda$.  Then let $\dC$ be the infinite dimensional
vector space over $\C$ generated by $d(l,\lambda)$ for all $l\in\N$
and all $\lambda\in\C$.  Let $\dC(\lambda)\subset\dC$ for
$\lambda\in\C$ be the subspace spanned by $d(l,\lambda)$ for all
$l\in\N$.  Stratify these subsets into
$\dC(l,\lambda)\subset\dC(\lambda)$ which is the subspace spanned by
$d(\alpha,\lambda)$ for $0\leq\alpha\leq l-1$.  (For convenience, we
define $\dC(0,\lambda)$ to be the empty set.)

Let $C$ be any $M$ dimensional subspace of $\dC$.  Here I will recall
the mapping which associates a point of $Gr_1$ to $C$ \cite{W}.  Let
$V_C$ be the vector space of polynomials in $z$ which satisfy the
condition $c(f)=0$ for each $c\in C$.  Finally, picking a monic
polynomial, $q(z)$, of degree $M$ the Hilbert closure of the set
$q^{-1}V_C$ in $H$ is a point $W\in Gr_1$.

This suggests the following definition.  Let $Gr(\dC)$ be the set of
finite dimensional linear subspaces of $\dC$.  Then letting $$
Gr_1^*=\left\{(C,q)\in Gr(\dC)\times \C[z]\bigg|\dim C=M\hbox{ and
}q=z^M+\sum_{i=0}^{M-1}c_iz^i\right\} $$ allows us to associate a
point $W\in Gr_1$ to each $W^*=(C,q)\in Gr_1^*$ by the {\it dual
mapping\/} described above.\footnote{%
Technically, the image of
certain points in $Gr_1^*$ are not contained in $Gr$ as formulated by
Segal and Wilson.  In particular, if $q(\zeta)=0$ for some
$\zeta\in\C$ such that $|\zeta|=1$, then the corresponding subspace
may not be contained in $L^2(S^1,z)$ unless it undergoes a scaling
transformation.  Without altering the results of the present paper,
one may resolve this problem either by restricting the definition of
$Gr_1^*$ to contain only points whose image is in $Gr$ or by extending
the definition of $Gr_1$ to contain subspaces which would be in $Gr$
after rescaling.  }%
Although this mapping is onto (since every
element of $Gr_1$ can be expressed in this way) it is not injective.
Note, for example, that $(\{d(0,0)\},z)$ and $(\{d(0,0),d(1,0)\},z^2)$
both get sent to the vacuum solution, $H_+\in Gr_1$.

The following lemma demonstrates that the choice of a polynomial $q$
only alters the gauge of the associated KP solution.  Consequently,
for most applications it will be sufficient to consider only the case
$q=z^M$ where $M$ is the dimension of $C$.  Only later when
bispectrality is being considered will gauge become significant.

\begin{lemma}{Varying the choice of $q$ merely affects the associated
solution by a gauge transformation.}  \end{lemma} \noindent{\bf
Proof:} Let $W_1^*=(C,q_1)$ and $W_2^*=(C,q_2)$ be two points of
$Gr_1^*$ with the same condition space $C$ and let $W_i\in Gr_1$ be
the image of $W_i^*$ under the dual mapping.  It is clear from the
definition of the mapping that the gauge transformation
$q_1/q_2\in\Gamma_-$ will take $W_1$ to $W_2$.  \qed

For any point of $Gr_{1}^*$, given by the condition space $C=\{c_i\}$
with basis $\{c_1,c_2,\ldots,c_M\}$ and $q=z^M$, it is simple to
calculate the corresponding $\tau$.  The definition of $\tau$ in
\cite{SW} is equivalent to $\tau=\det\M$ \cite{W}, where $\M$ is the
$M\times M$ matrix $$ \M_{ij}=c_i(z^j e^{\sum t_i z^i}).  $$ (This
determinant can be viewed as the Wronskian of the $M$ one condition
solutions given by the individual $c_i$'s.)  Notice that another
choice of basis for $C$ only affects $\tau$ by a constant multiple.

\subsection{The Finite Grassmannians}

Recall that the grassmannian $Gr(M,N)$, which is made up of $M$
dimensional subspaces of an $N$ dimensional vector space $V$, is
isomorphic to the grassmannian $Gr(N-M,N)$.  This isomorphism follows
from the Principle of Duality \cite{HP}, since such an isomorphism is
given by sending $W\in Gr(M,N)$ to the $N-M$ dimensional subspace
$W^*\subset V^*$ such that $w^*(w)=0$ for all $w\in W$ and $w^*\in
W^*$.  Consequently, $Gr(N-M,N)$ is referred to as the dual
grassmannian of $Gr(M,N)$.  This section will demonstrate that the
$\Gamma_+$ orbit of any point $W^*$ is contained in a finite
dimensional sub-grassmannian of $Gr_1^*$.  These finite dimensional
grassmannians are actually the dual grassmannians of the
sub-grassmannians of $Gr_1$ which are their image under the dual
mapping (thereby justifying the terminology).  In addition, it will be
seen that they decompose into a disjoint union of $\Gamma_+$ orbits.

Let $\mu:\C\to\N$ be called a {\it singularity bounding function\/}
(or simply bounding function) if the set $\C-\mu^{-1}(0)$ is a finite
set of points.  This finite set on which $\mu$ is non-zero is called
the support of $\mu$ ($\supp \mu$).  Define the dimension of $\mu$ by
$\dim\mu=\sum_{\lambda\in\C}\mu(\lambda)$.

Given a bounding function, $\mu$, let $$
\dC(\mu)=\bigoplus_{\lambda\in\C} \dC(\mu(\lambda),\lambda).  $$ Then
$\dC(\mu)$ is a $\dim\mu$ dimensional vector space.  For any
$M<\dim\mu$, the set of $M$ dimensional subspaces of $\dC(\mu)$ is
then a well defined grassmannian: $Gr(M,\dC(\mu))$.  In order to
consider the images under the dual mapping of points in this
grassmannian we must specify a choice of $q$.  We will denote this by
$Gr_q(M,\dC(\mu))$ and assume that $q=z^M$ when none is specified.
The dual mapping from $Gr_q(M,\dC(\mu))$ to $Gr_1$ is injective.  (In
fact, once the choice of $M$ and $q$ is fixed, the dual mapping is
injective.)

Given a point $W^*=(C,z^M)\in Gr(M,\dC(\mu))$, consider the form of
the associated point $W\in Gr_{1}$.  Note first that the polynomial
$p(n,z)=z^n\prod(z-\lambda)^{\mu(\lambda)}$ will be in $V_C$ for any
$n\in\N$.  Thus, $ p(0,z)H_+\subset W\subset z^{-M}H_+$ demonstrates
that $W\in Gr_1$ according to the definition in \cite{SW}.  Then $W$
will have this form: $$
W=z^{-M}\{\omega_1,\omega_2,\ldots,\omega_N,p(0,z),p(1,z),\ldots\}, $$
in which everything is fixed apart from the choice of $\omega_i$.  The
$\omega_i$ can be any basis of the $N$ dimensional space of
polynomials of degree less than $\dim\mu$ which satisfy
$c(\omega_i)=0$ for all $c\in C$.  Since $M$ independent conditions
will have been applied to get this set, we know that $N=\dim\mu-M$.
Then, the dual mapping is an isomorphism from $Gr(M,\dC(\mu))$ to the
$Gr(\dim\mu-M,\dim\mu)$ which is its image in $Gr_1$.  In particular,
the two finite dimensional grassmannians are dual to each other and
$W$ is the space ``perpendicular'' to $W^*$ under the inner-product
defined by application of the condition to the function.  The
grassmannians corresponding to different choices of $q$ are trivially
isomorphic and all statements above apply to them as well.

\noindent{\bf Note:} One may coordinatize the image in $Gr_1$ with the
Pl\"ucker coordinates of a $Gr(M,\dim\mu)$ given by the subspace
spanned by $\{\omega_i\}$ in the vector space spanned by $\{z^i|0\leq
i \leq \dim\mu-1\}$.  It is then possible to choose a set of functions
for which these finite dimensional Pl\"ucker coordinates are
represented as the coefficients in a series summing to $\tau$.  Denote
by $S_{\mu}^q$ the finite set of $\tau$-functions which correspond to
the points in the image with only one non-zero Pl\"ucker coordinate.
In a method analogous to the infinite dimensional case, the
$\tau$-function of an arbitrary point of $Gr_q(M,\dC(\mu))$ can then
be written as the finite sum of these functions with the corresponding
Pl\"ucker coordinates as coefficients.  In particular, all
$\tau$-functions from this grassmannian are contained in the finite
dimensional space spanned by $S_{\mu}^q$.  If we write $\tau$ as an
arbitrary sum of elements from $S_{\mu}^q$, then the equations of the
KP hierarchy will act as Pl\"ucker relations for the coefficients.

\noindent {bf Example:} Consider the grassmannian $Gr(2,\dC(\mu))$
where $\mu(n)=3\delta_{0n}+\delta_{1n}$ and $q=z^2$.  The Pl\"ucker
coordinates of the grassmannian will be a set of six coordinates
$\pi_{ij}$ with $0\leq i < j \leq 3$.  Denote by $\tau_{ij}$ the
$\tau$-function corresponding to the point where only $\pi_{ij}$ is
non-zero.  These six $\tau$-functions are as follows: $$ \tau_{01} =
x^2+2x-2y+(x^2+2y-2x)e^{\sum
t_i}\qquad\tau_{03}=x^2-2y\qquad\tau_{13}=x $$ $$
\tau_{02}=-2x^2+4y+2-2(1-x)e^{\sum t_i} \qquad \tau_{12}=-2x-1+e^{\sum
t_i}\qquad \tau_{23}=1 $$ Notice that three of these functions are
{\it not\/} the Schur polynomials which arise in the standard
representation.  Every $\tau$-function from $Gr(2,\dC(\mu))$ is
contained in the vector space spanned by the $\tau_{ij}$, and can be
written as $$ \tau=\sum_{0\leq i<j \leq 3} \pi_{ij}\tau_{ij}, $$
although an arbitrary sum of this form is not necessarily a
$\tau$-function.  Inserting this arbitrary sum into the Hirota
bilinear form of the KP equation yields $$ 48e^{\sum
t_i}(\pi_{03}\pi_{12}-\pi_{02}\pi_{13}+\pi_{12}\pi_{13})=0 $$ which,
apart from the exponential coefficient, is the Pl\"ucker relation for
the coordinates of $Gr(2,4)$.

\begin{definition}\label{gamma+} Define the action of
$g(z)=e^{cz^n}\in\Gamma_+$ on $d(l,\lambda)\in\dC(\mu)$ by the formula
$$ g(d(0,\lambda)) = e^{-c\lambda^n}d(0,\lambda) \qquad
g(d(l,\lambda)) = e^{-c\lambda^n}\left(d(l,\lambda)-\sum_{i=1}^l
k_{l-i}d(l-i,\lambda)\right),\hbox{ for } l\geq1 $$ with the $k_{l-i}$
for $0\leq i \leq l$ defined recursively as $$
k_{l-1}=l\frac{{\partial}}{{\partial}z}\log
g(z)\bigg|_{z=\lambda}\qquad k_{l-i}=\frac{1}{g}\left({l\choose
i}g^{(i)}-\sum_{j=1}^{i-1}k_{l-j}{{l-i}
\choose{i-j}}g^{(i-j)}\right)\bigg|_{z=\lambda}
$$ Extend this linearly to define the action of the group $\Gamma_+$
on $\dC(\mu)$.  \end{definition}

Since $g(c)$ has the same support and bound as the condition $c$,
$\Gamma_+$ acts on the linear space $\dC(\mu)$.  Furthermore, since
$\Gamma_+$ acts linearly and has no kernel, it takes one $M$
dimensional subspace to another. Consequently, $\Gamma_+$ acts on the
finite dimensional grassmannian $Gr_q(M,\dC(\mu))$ under
Definition~\ref{gamma+}.

Definition~\ref{gamma+} was chosen so that for any $c\in\dC(\mu)$,
$g\in\Gamma_+$ and $f\in\C\iseries{z}$ $$ c(f)=0 \hbox{ iff }
g(c)(g\cdot f)=0.  $$ Therefore, this definition of the action of
$\Gamma_+$ on $Gr_1^*$ coincides with the multiplicative action on
$Gr_1$ via the dual mapping.

\subsubsection{Grassmannians and Generalized Jacobians}

The $\Gamma_+$ orbit of any point $W$ in $Gr$ which is ``rank one
algebro-geometric'', modulo the action of $\Gamma_-$, will be the
generalized jacobian of the spectral curve.  In fact, in the case of
$Gr_1$, taking the quotient by $\Gamma_-$ is unnecessary and we have
the following result:

\newtheorem{proposition}[theorem]{Proposition}
\begin{proposition}\label{orb/jac}The $\Gamma_+$ orbit of a point
$W\in Gr_1$ is isomorphic to the (generalized) jacobian of the
corresponding spectral curve.\end{proposition}

\noindent Proposition~\ref{orb/jac} follows immediately from
Lemma~\ref{+/-}.

\begin{lemma}\label{+/-}If $\calg\not=1\in\Gamma_-$ is a non-trivial
gauge transformation and $W\in Gr_1$, then $\calg W$ is not in the
$\Gamma_+$ orbit of $W$.  \end{lemma} \noindent{\bf Proof:} By the
formula for $\tau$ as the determinant of the Wronskian matrix $\M$,
one can write $\tau_W$ as $$ \tau_W=\sum_{j=1}^N p_j(\t) e^{\sum
a_{ij} t_i} $$ where the $p_j$ are of bounded degree in all the $t_i$.
Then, define $\omega(\tau_W,i)$ to be the unordered $N$-tuple
$\{a_{ij}|0\leq j\leq N\}$, which is well defined despite the
projective ambiguity in $\tau$.  Note that the $\omega(\tau_W,i)$ are
preserved under $\Gamma_+$ translation of $W$.  Now suppose that
$\calg\in\Gamma_-$ is a non-trivial gauge transformation.  Then,
$\tau_{\calg W}=\hat\calg\tau_W$ where $$\hat\calg=e^{\sum \alpha_i
t_i}$$ with not all $\alpha_i=0$.  Let $i_0$ be such that
$\alpha_{i_0}\not=0$ and note that $\omega(\tau_{\calg
W},i_0)\not=\omega(\tau_W,i_0)$ since every element has undergone and
additive shift of $\alpha_{i_0}$.  Therefore, $\calg
W\not\in\Gamma_+W$.  \qed

Therefore, the $\Gamma_+$ orbit of any point $W^*\in Gr_q(M,\dC(\mu))$
will be the (generalized) jacobian of a (singular) rational curve.
Furthermore, it is clear that these orbits are disjoint sets.  Thus,
one may conclude that any $Gr(M,\dim\mu)$ can be written as the
disjoint union of the jacobians of all rational curves which can be
described by $M$ differential conditions bounded by $\mu$.  Since any
$N\in\N$ is realizable as $\dim\mu$ for many different choices of
$\mu$ (essentially determined by a partition of $N$), one can actually
say:

\begin{proposition}\label{gr/jac}For any $M\leq N\in\N$, the
grassmannian $Gr(M,N)$, can be decomposed into the disjoint union of
the generalized jacobians of all curves represented by $M$ conditions
bounded by $\mu$ for any $\mu$ of dimension $N$.\end{proposition}

The two decompositions of $Gr(1,2)$, corresponding to the two
partitions of the number 2 serve as an enlightening example.  Consider
first the bounding function $\mu_a(n)=2\delta_{an}$ and $M=1$.  A
point of the corresponding grassmannian is determined by a condition
$c_1f'(a)+c_2f(a)=0$ for $(c_1,c_2)\in \P^1\C$.  There are two
distinct $\Gamma_+$ orbits which form this set.  In the case $c_1=0$,
we have the trivial solution $u=0$ which is fixed by all $\Gamma_+$
action.  Alternatively, if $c_1\not=0$, let $c=c_2/c_1$ and the
condition can be written simply as $f'(a)+cf(a)=0$.  One may check
that the subgroup of $\Gamma_+$ generated by elements of the form
$e^{z^i-ia^{i-1}z}$ is the stabilizer of this condition (for any value
of $c$).  The action of $\Gamma_+$ can thus be understood by taking
the quotient by this stabilizer, any element of which has a
representative of the form $e^{\theta z}$.  Note that this element
takes the condition above to $f'(a)+(c-\theta)f(a)=0$.  In particular,
the entire orbit is isomorphic to $\C$, which is indeed the
generalized jacobian of a rational curve with a simple cusp at $a$.

Alternatively, consider the grassmannian corresponding to
$\mu_{ab}(n)=\delta_{an}+\delta_{bn}$ and $M=1$.  A point of this
grassmannian is specified by a condition $c_1f(a)+c_2f(b)=0$ for
$(c_1,c_2)\in\P^1\C$ as before.  Note, however, that the case in which
either $c_i$ is zero corresponds to the trivial solution, and thus to
two distinct $\Gamma_+$ orbits.  We are then left with the case
$c_1,c_2\in\C^{\times}=\C-\{0\}$.  The condition $f(a)+cf(b)=0$ (where
$c=c_2/c_1\not=0$) is stabilized by any element of the form $$
exp\left({z^i-\frac{b^i-a^i}{b-a}z}\right).  $$ Once again, we may now
consider any element of $\Gamma_+$ to be represented by an element of
the form $e^{\theta z}$.  (In this case, we have not taken the
quotient by the entire stabilizer since there remains a periodicity in
$\theta$.)  The action of this element on the condition gives
$f(a)+e^{\theta(b-a)}cf(b)=0$.  In particular, we can translate $c$ to
any value in $\C^{\times}$, which is the generalized jacobian of a
rational curve with a node given by identifying the points $a$ and
$b$.

\noindent{\bf Note:} A clever argument of G. Segal (related to me by
E. Previato) demonstrates that the result of Lemma~\ref{+/-} is not
necessarily true outside of $Gr_1$.  More precisely, if any
non-trivial gauge transformation takes $W$ outside its $\Gamma_+$
orbit, then $A_W\subset\C[z]$.  Thus, in general, we should not expect
that the KP orbit of the operator $L_W$ is isomorphic to the
$\Gamma_+$ orbit of $W$.

\subsection{Rational Solutions}\label{ratsols}%

Given a point $(C,q)\in Gr_{1}^*$, the corresponding solution will be
rational if and only if $C$ is spanned by $\{c_1,c_2,\ldots,c_M\}$
where $c_i\in \dC(\lambda_i)$ for some $\lambda_i\in\C$.  (That is,
each condition must involve only one point.  This limits the spectral
curves to rational curves with cusps.)  This is proved in \cite{W},
but the basic idea can be understood by studying the matrix $\M$ which
corresponds to the gauge $q=z^M$.  Note that it is only when $C$ has
this property that the $i^{\scriptsize{th}}$ column is made up of
polynomials multiplied by $e^{\sum \lambda_i^j t_j}$.  Then, using
elementary properties of the determinant, one can factor out these
exponentials and find that $\tau$ is the determinant of a matrix of
polynomials multiplied by the exponential of a linear function of $x$.
Consequently, the general form of such a $\tau$ is easily seen to be:
$$
(\hbox{*})\qquad\tau=p(\t)\prod_{i=1}^M\left(exp\left(\sum_{j=1}^{\infty}
\lambda_i^j t_j\right)\right), $$ where $p(\t)$ is a polynomial of
degree $N$ in $x$ and a polynomial of degree less than or equal to $N$
in each of the other time variables, $t_i$, and $\lambda_i$ are the
singular points.  This clearly makes
$u=2\frac{{\partial}^2}{{\partial}x^2}\log\tau$ into a rational
function.


\subsection{Translating Krichever's Parameters}

The function $q$ plays a fundamentally different role in the two
methods discussed for generating rational solutions.  In the dual
grassmannian, the choice of $q$ merely determines the gauge.  In
Krichever's method, however, the gauge is fixed by the form $(Kr)$ of
the wave function.  Thus, once the $\lambda_i$ have been fixed, the
choice of $q$ is the only way of determining the solution.  The
difference arises from the fact that Krichever's condition is applied
to the function $\psi$, which is an element of $q^{-1}\overline{V_C}$
rather than an element of $\overline{V_C}$ itself.  Then the ``chain
rule'' gives us the ability to relate the two sets of parameters.

\begin{claim}\label{trans}If we choose the polynomial $q$ and
$c_i(f(z))=f'(z)-\frac{q'(\lambda_i)}{q(\lambda_i)}f(z)|_{z=\lambda_i}$
as our conditions in the dual grassmannian, this will lead to the same
solution as choosing $q$ and $\{\lambda_i\}$ in Krichever's method.
\end{claim}

\noindent {\bf Proof:} \begin{eqnarray*} \M_{ij} &=&
c_i(z^{j-1}e^{\sum t_nz^n}) = \left(z^{j-1}e^{\sum
t_nz^n}\right)'-\frac{q'(\lambda_i)}{q(\lambda_i)}e^{\sum t_n
z^n}\bigg|_{z=\lambda_i}\\ &=&
q(\lambda_i)\frac{{\partial}}{{\partial}z}z^{j-1}\frac{e^{\sum
t_nz^n}}{q(z)}\bigg|_{z=\lambda_i}
 = \Theta_{ij}q(\lambda_i)e^{\sum \lambda_i^n t_n} \end{eqnarray*}

The presence of the factor $q(\lambda_i)$ merely affects the
determinant by a constant multiple.  Furthermore, in \cite{Kr},
Krichever has removed the exponential part of each $\theta_{ij}$.  (As
we shall see below, this latter difference corresponds to a gauge
transformation.)  To complete the proof it is sufficient to see that
these differences will not alter the associated rational solution.
\qed

\noindent{\bf Note:} It is known that the gauge of Krichever's wave
function is identified by the property that $1\in W$.  This can be
determined, for example, from the results of \cite{SW}.  This
statement is also easily observed from Claim~\ref{trans} since $q(z)$
clearly satisfies the conditions.

\section{Bispectrality}\label{sec:bisp}%

Given a wave function $\psi(x,z)$, which is an eigenfunction in a
generalized Schr\"odinger equation with spectral parameter $z$: $$
L(x,\frac{{\partial}}{{\partial}x})\psi=\Theta(z)\psi $$ we say that
$\psi$ is {\it bispectral\/} if it also satisfies an analogous
equation with the roles of $x$ and $z$ reversed: $$
Q(z,\frac{{\partial}}{{\partial}z})\psi=\Phi(x)\psi.  $$ This property
was first discussed in \cite{DG} in connection with some questions
arising in medical imaging.  Surprisingly, it was found that the wave
functions corresponding to rational KdV solutions were bispectral.
Zubelli \cite{Z} extended this result to all the rational KP solutions
from $Gr_0$ by explicitly constructing operators in the spectral
parameter which demonstrate the bispectrality of the corresponding
wave functions.

Wilson \cite{W} completely classified the bispectral wave functions of
the KP hierarchy which correspond to a rank one ring of commuting
differential operators.  (That is, \cite{W} determined all bispectral
wave functions which can be constructed from an isospectral flow of
{\it line\/} bundles.)  He first demonstrated that such wave functions
must correspond to (vanishing or non-vanishing) rational
solutions. Moreover, he concluded that a {\it unique\/} point in the
$\Gamma_-$ orbit of each rational solution of $Gr_1$ corresponds to a
bispectral wave function. In particular, it was shown that given a
condition space $C=\{c_i\}$ such that $c_i$ is a differential
condition at the point $\lambda_i$, the unique bispectral point
corresponding to this condition space is $q^{-1}\overline{V_C}$ for
$q(z)=\prod (z-\lambda_i)$.  I will refer to the point corresponding
to this choice of $q$ as the {\it bispectral gauge\/}.

Furthermore, it was shown that the stationary wave function
corresponding to the bispectral gauge, $\psi(x,z)$, remains a wave
function of $Gr$ in bispectral gauge when $x$ and $z$ are
interchanged\footnote{ It is clear that if $\psi(z,x)$ is also a wave
function then $\psi(x,z)$ is bispectral.  To see the more powerful
fact that these are the only bispectral wave functions in $Gr_1$,
please refer to the proof in \cite{W}.  }.  It is in this context that
the {\it bispectral involution\/} arises naturally.  Consider the
involution, $\beta$, acting on functions of $x$ and $z$ which switches
the variables (i.e., $\beta(f(x,z))=f(z,x)$).  Then, as a result of
\cite{W}, $\beta$ is an involution on bispectral wave functions
corresponding to rational KP solutions.

If we write $\psi(x,z)=\hat\psi e^{xz}$, then
$\hat\psi=1+\alpha_1(x)z^{-1}+\alpha_2(x)z^{-2}+\cdots$.  It is
therefore clear that $\lim_{z\to\infty}\hat\psi=1$.  Note also, if
$\calg\not=1\in\Gamma_-$ and $\lim_{x\to\infty}\hat\psi=1$ then $$
\lim_{x\to\infty}\calg\hat\psi=\calg\not=1.  $$ From these facts it is
simple to deduce the following corollary.

\begin{corollary}\label{limx}{A stationary wave function
$\psi=\hat\psi e^{xz}$ of a rational solution to the KP hierarchy
satisfies $$ \lim_{x\to\infty}\hat\psi=1 $$ if and only if $\psi$ is
in the bispectral gauge.}  \end{corollary}

\subsection{Tau Functions of Bounded
Degree} \begin{lemma}\label{bdd}{The $\tau$ function of a rational
solution in the bispectral gauge is of {\it bounded degree\/} in each
time variable.  (That is, there is one $N$ such that $\tau$ is a
polynomial of degree at most $N$ in each time variable.)}  \end{lemma}
\noindent{Proof:} Recall from Section~\ref{ratsols} that the $\tau$
corresponding to a rational solution in $Gr_{1}^*$ with $q=z^M$ is in
the form $(\hbox{*})$.  It will be demonstrated that the element of
$\Gamma_-$ which will take this rational solution to its bispectral
gauge results in a $\tau$ of bounded degree.

The action of the gauge group $\Gamma_-$ on the representation $\tau$
can be expressed as follows.  Let $\calg=e^{\alpha z^{-i}}\in\Gamma_-$
and $\hat\calg=e^{-\alpha i t_i}$ and denote by $\calg W$ the
translation of $W\in Gr$ by $\calg$.  Then we have: $$ \tau_{\calg
W}=\hat\calg\tau_W.  $$ It is then simple to see that multiplication
by the gauge transformation $$ \calg(a)=1-az^{-1} =e^{\ln 1-az^{-1}} =
e^{-\sum(a^i/i) z^{-i}} $$ corresponds to multiplying $\tau$ by
$e^{\sum a^i t_i}$.

To move a point with $q=z^M$ to its bispectral gauge, we would
multiply by $$
\frac{z^M}{\prod(z-\lambda_i)}=
\prod\frac{z}{z-\lambda_i}=\prod\frac{1}{1-\lambda_iz^{-1}}=\left(\prod
\calg(\lambda_i)\right)^{-1}.  $$ Note that the result of this
transformation on a $\tau$ in the form $(*)$ is the elimination of the
exponential part, leaving $\tau=p(\t)$.  \qed

\begin{theorem}\label{thm:bisp}{The following are equivalent
conditions on $W\in Gr$: \begin{enumerate} \item There exists an
$N\in\N$ such that $\tau_W$ is a polynomial of degree at most $N$ in
each of the variables $t_i$.

\item $u=2\frac{\partial^2}{\partial x^2}\log\tau_W$ is a vanishing
rational solution and $\psi_W(x,z)=\psi_{W'}(z,x)$ for some $W'$ with
the same properties.

\item $W=q^{-1}\overline{V_C}$ where $V_C$ is the set of polynomials
in $z$ satisfying the conditions $\{c_1,\ldots,c_m\}$, the condition
$c_i$ involves only derivatives evaluated at the point $\lambda_i$,
and $q=\prod(z-\lambda_i)$.

\item $\tau_W$ is a polynomial in $x$ and the coefficient of the
highest degree term of $x$ is constant.\footnote{%
T. Shiota
\cite{Taka} has recently completed a study of the $\tau$-functions of
the KP equation which are monic polynomials in $x$ and their extension
to the KP hierarchy.  } \end{enumerate}} \end{theorem} \noindent{\bf
Proof:} {\bf (1) implies (2):} A $\tau$ of bounded degree is
necessarily a polynomial in $x$, and thus
$u=2\frac{{\partial}^2}{{\partial}x^2}\log\tau$ is a (vanishing)
rational solution.  Given that \cite{Kr} finds all such solutions as
resulting from cuspidal rational curves, one may conclude that $W$
comes from Wilson's construction for rational curves with cusps.
Furthermore, it was shown in Lemma~\ref{bdd} that one only gets a
$\tau$ of bounded degree in this construction with the bispectral
choice of $q$.  Consequently, we also have that $W\to W'$ under the
bispectral involution.

{\bf (2) implies (3):} That (2) and (3) are equivalent is a major
result of \cite{W}.

{\bf (3) implies (1):} This is what I have shown above.

Thus we have that (1), (2) and (3) are equivalent.

That (4) is equivalent to (1--3) is most simply seen by comparing it
to (2).  Since $\tau_W$ is a polynomial in $x$, the corresponding
solution is clearly a vanishing rational solution.  Furthermore, the
fact that the coefficient of the highest degree term in $x$ is
constant is equivalent to the fact that $\lim_{x\to\infty}
\hat\psi=1$.  (If $\tau$ is a polynomial in $t_i$ and $$
\hat\psi=1+\frac{h(\t,z)}{\tau(\t)q(z)} $$ then $h$ is of lower degree
than $\tau$ in $t_i$ if and only if the coefficient of the highest
power of $t_i$ in $\tau$ is constant.  This can be determined from the
formula relating $\tau$ and $\psi$ \cite{SW}.)  So, (4) is really only
a restatement of (2).  \qed

Since every rational solution has a $\tau$ of bounded degree, this can
be seen as an extension of Corollary~\ref{polyxyt} to the arbitrary
case and all of the time variables.  Furthermore, from the fact that
(4) implies (1), we get the following corollary.  \begin{corollary}{If
$\tau_W$ is polynomial in $x$ with constant coefficient on the highest
degree term, then $\tau_W$ is polynomial in each $t_i$.}
\end{corollary}

\section{The Bispectral Involution}\label{sec:beta}%

The correspondence between rational KP solutions and Calogero-Moser
particle systems is simple to describe in the case that the solution
is in the bispectral gauge.  In that case, the positions of the
particles $x_j$ are the roots of the polynomial $\tau_W$ in the
variable $x$ and the momenta $y_j$ are the instantaneous velocity of
the $x_j$'s under the second time flow.  By using this correspondence,
it is possible to consider the bispectral involution as a map on
particle systems.  It can then be compared to the well known
linearizing map $\sigma$ \cite{AMcM} which is also an involution on
Calogero-Moser particle systems.  The results are summarized by the
following theorems and proved below.

In the case that a KP solution can be represented by conditions of the
form $ f'(\lambda_i)+\gamma_i f(\lambda_i)=0$ for distinct
$\lambda_i$, the parameters $\lambda_i$ and $\gamma_i$ will be
referred to as the {\it dual grassmannian coordinates\/} of the
solution.  The bispectral involution demonstrates a symmetry between
the dual grassmannian coordinates of a solution and the corresponding
\cmps.  This can be applied to determine the action of $\beta$ on
\cmpss\ in terms of dual grassmannian coordinates or the action on
dual grassmannian coordinates in terms of \cmpss\ as shown in the
following two theorems:

\begin{theorem}\label{image}{If the \cmps\ $(\vec{x},\vec{y})$ is
determined by the dual grassmannian parameters
$(\vec{\lambda},\vec{\gamma})$ ($\lambda_i\not=\lambda_j$ for
$i\not=j$), then $\beta(\vec{x},\vec{y})=(\vec{\BISxi},\vec{\BISeta})$
is given by $$\BISxi_i=\lambda_i$$ and
$$\BISeta_i=\gamma_i+\sum_{j\not=i}\frac{1}{\lambda_i-\lambda_j}.$$}
\end{theorem}

\begin{theorem}\label{coords}{Given a bispectral rational KP solution
$W^*\in Gr_1^*$ that corresponds to a \cmps\ $(\vec{x},\vec{y})$ such
that the $x_j$ are distinct, the condition space with basis: $$
c_i(f)=f(x_i)+\left(y_i-\sum_{j\not=i}\frac{1}{x_i-x_j}\right)f'(x_i)=0
$$ and the polynomial $q(z)=\prod(z-x_i)$ are the point $\beta(W^*)\in
Gr_1^*$.}  \end{theorem}

Then, by considering the composition of the bispectral involution with
the motion of the particle systems, it is determined that:

\begin{theorem}\label{action-angle}{$\beta$ acting on \cmpss\ with
distinct particle positions is a linearizing map in the sense that it
fixes the positions and linearizes the momenta under composition with
time flows.}  \end{theorem}

Thus, $\beta$, which is an involution by construction, has the
surprising property of being a linearizing map.  Alternatively, the
map $\sigma$ which is a linearizing map by construction, is
mysteriously an involution.  Finally, it is noted that the two
involutions are basically the same.

\begin{theorem}\label{sigma}{Given a \cmps\
$(\vec{x}(\ttwo),\vec{y}(\ttwo))$, $$
\sigma(\vec{x}(\ttwo),\vec{y}(\ttwo))=
(\vec{\Mxi},2\vec{\Mxi}\ttwo+\vec{\Meta})
$$ and $$
\beta(\vec{x}(\ttwo),\vec{y}(\ttwo))=
(\vec{\BISxi},2\vec{\BISxi}\ttwo+\vec{\BISeta})
$$ for constant $\vec{\Mxi}$, $\vec{\Meta}$, $\vec{\BISxi}$ and
$\vec{\BISeta}$.  Furthermore, they are related by the fact that
$\vec{\BISxi}=-\vec{\Mxi}$.}  \end{theorem}


\subsection{Bispectral Flow}\label{sec:bisp-flow}%

Let $C$ be an $N$ dimensional space of conditions with basis
$\{c_1,\ldots,c_N\}$ such that $c_i\in \dC(\lambda_i)$ and let
$q=\prod(z-\lambda_i)$.  Denote by $W^*$ the point $(C,q)\in Gr_1^*$
and its image in $Gr_1$ by $W$.  The spectral curve corresponding to
this KP solution is a rational curve with cusps at the points
$\lambda_i$ and its stationary wave function $\stat$ is in the form
$(B)$ with poles in $z$ at the cusps of the spectral curve and poles
in $x$ at the zeroes of the polynomial $\tau_W(x,0,0,\ldots)$.  These
values of $x$ are those for which the corresponding point of the $t_1$
orbit of $W$ leaves the ``big cell'' of $Gr$ \cite{SW}.

The bispectrality of $\stat$ implies that its image under the
bispectral involution, $$\beta(\stat(x,z))=\stat(z,x), $$ is the wave
function of another bispectral rational solution.  Since
$\beta(\stat)$ exchanges the roles of $x$ and $z$, it is clear from
the observations in the previous paragraph that the spectral curve of
the KP solution with wave function $\beta(\stat)$ has cusps at those
points $z$ for which $\tau_W(z,0,0\ldots)=0$.

Let $W(t_i)$ be the time dependence of the point $W\in Gr_1$ under the
$i^{\,\hbox{\scriptsize th}}$ KP flow and $\beta(W(t_i))$ its image
under the bispectral involution.  Recall that the zeroes of the
polynomial $\tau_W(x,0,0,\ldots)$ move as a Calogero-Moser system of
particles under the KP flow.  Then the composition of any KP flow with
the bispectral involution, the {\it bispectral flow\/}, is seen to be
a non-isospectral flow for which the cusps of the corresponding
spectral curve behave as a Calogero-Moser system.  In the case of the
first KP flow given by $x\to x+c$, the induced bispectral flow would
clearly be given by a linear deformation of the spectral parameter
$z$, which is an example of a Virasoro flow \cite{Vir}.  As will be
demonstrated below, it was essentially a bispectral flow which was
utilized in \cite{AMcM} as a linearized flow of the Calogero-Moser
particle system.

\subsection{Calogero-Moser Particle Systems}

The phase space of an $n$-particle \cmps\ is given by $2n$ complex
numbers, $x_i$ and $y_i$, $1\leq i \leq n$, $x_i\not=x_j$ for
$i\not=j$, where $\vec{x}=(x_1,\ldots,x_n)$ represent the positions of
$n$ particles and $\vec{y}=(y_1\ldots,y_n)$ represent their
instantaneous velocity.  Here we will be mainly interested in their
motion under the second Hamiltonian: $H=\sum y_i^2 -\sum
{(x_i-x_j)^{-2}}$.  Airault, McKean and Moser \cite{AMcM} introduced
the linearizing map
$\sigma(\vec{x},\vec{y})\to(\vec{\Mxi},\vec{\Meta})$ in the case when
all the phase space variables are real, where $\Mxi_i$ is the
asymptotic velocity of the $i^{th}$ particle and $\Meta_i$ is an
asymptotic relative position.  It is clear, by physical consideration,
that $\Mxi_i$ is a constant of motion and that $\Meta_i$ is a linear
function of the time variable.  It was unexpected, however, that the
map $\sigma$ would be an involution.  Yet, it was shown in
\cite{AMcM}, that $\sigma(\vec{\Mxi},\vec{\Meta})=(\vec{x},\vec{y})$.
It is important to note that the matrix
$$\Lambda_{ij}=y_{i}\delta_{ij}+\frac{1-\delta_{ij}}{x_i-x_j}$$ has
$\{\Mxi_i\}$ as its eigenvalues and that the analogous matrix
$\Lambda^{\sigma}$ written in terms of $\Mxi_i$ and $\Meta_i$ has
$\{x_i\}$ as its eigenvalues.  Therefore, one can view $\sigma$ as a
map which exchanges spacial and spectral values.  This, superficially,
indicates a relationship to the bispectral involution.

\noindent{\bf Note:} Although vector notation is being used, particle
systems within this paper are unordered sets of pairs $(x_i,y_i)$.
Thus, two particle systems which differ only by a permutation of the
index $i$ are considered to be the same particle system.

\subsection{Dual Grassmannian Coordinates of Krichever's Solutions and
the Associated Particle System}

It is clear from Theorem~\ref{trans} that the rational solutions that
are constructed in Theorem~\ref{Kr2} are also given by the dual
grassmannian construction with a choice of $2n$ parameters:
$\lambda_i$ ($\lambda_i\not=\lambda_j$ if $i\not=j$) and $\gamma_i$
for $1\leq i \leq n$, where the point $W^*$ has a condition space
spanned by $n$ differential conditions: $$
c_i(f)=f'(\lambda_i)+\gamma_if(\lambda_i)=0$$ and can be placed in the
bispectral gauge by the choice of polynomial
$q(z)=\prod(z-\lambda_i)$.  In the case that a KP solution can be
represented by a condition space of this form with distinct singular
points $\lambda_i$, the parameters $\lambda_i$ and $\gamma_i$ will be
referred to as the {\it dual grassmannian coordinates\/} of the
solution.  The dual grassmannian coordinates will be written using the
same notation as a particle system: $(\vec{\lambda},\vec{\gamma})$.

In this section, we will only be concerned with this solution and its
motion under the second time flow.  We therefore suppress all time
variables other than $x$ and $\ttwo$.  Then the time dependent wave
function of the solution described above can be written in the form
\begin{eqnarray*} \psi(x,\ttwo,z) &=&
\left(1+\frac{p(x,\ttwo,z)}{\prod(x-x_i(\ttwo))\prod(z-\lambda_i)}\right)e^{xz+\ttwo
z^2}\\
 &=&
\left(1+\frac{1}{\prod(z-\lambda_i)}\sum\frac{a_i(\ttwo,z)}{x-x_i(\ttwo)}\right)e^{xz+\ttwo
z^2 }\\
 &=&
\left(1+\frac{1}{\prod(x-x_i(\ttwo))}\sum\frac{\alpha_i(x,\ttwo)}{z-\lambda_i}\right)e^{xz+\ttwo
z^2}\\ \end{eqnarray*} where $p(x,\ttwo,z)$ is a polynomial of degree
less than $n$ each variable and the residues $a_i$ and $\alpha_i$ are
merely determined by a partial fractions expansion in $x$ and $z$
respectively.  The stationary wave function $\psi(x,z)$ is merely
$\psi(x,0,z)$, i.e. the wave function at $\ttwo=0$.

Notice that the positions of the poles of $\psi$ in the variable $x$,
which are given by the functions $x_i$, move in time whereas the
positions of the poles in $z$ are fixed.  As is often the case in the
study of rational solutions to integrable equations, the motion of the
poles in $x$ are equivalent to an integrable Hamiltonian system of $n$
particles.  Let $y_i=\frac{1}{2}\frac{d}{d\ttwo}x_i|_{\ttwo=0}$ be the
``instantaneous velocity'' of the particle at position $x_i$.  Thus we
associate the particle system $(\vec{x},\vec{y})$ to the wave function
$\psi$.

Denote by $\CM_0$ the subset of the phase space of an $n$ particle
\cmps\ for which the positions, $x_i$, are distinct.  If the dual
grassmannian parameters $\lambda_i$ and $\gamma_i$ were chosen such
that $(\vec{x},\vec{y})\in\CM_0$ let $\Lambda$ be the Moser matrix $$
\Lambda_{ij}=y_i \delta_{ij} + \frac{1-\delta_{ij}}{x_i-x_j}.  $$ Then
the motion of the $x_i$ under the $\ttwo$ flow is a \cmps, given by
the Hamiltonian $H=\hbox{tr}\,\Lambda^2$.

\subsection{The Particle System of the Bispectral Dual of the Wave
Function}

Recall that in the bispectral gauge, the stationary wave function
remains a stationary wave function of the KP hierarchy if the
variables $x$ and $z$ are interchanged \cite{W}.  So, in particular,
we may talk about the bispectral dual of the wave function
\begin{eqnarray*} \psi^{\beta}(x,z) &=& \psi(z,x)\\
 &=&
\left(1+\frac{1}{\prod(x-\lambda_i)}\sum\frac{a_i(x)}{z-x_i}\right)e^{xz
}\\
 &=&
\left(1+\frac{1}{\prod(z-x_i)}\sum
\frac{\alpha_i(z)}{x-\lambda_i}\right)e^{xz}.  \end{eqnarray*}

There are two reasonable ways to consider the time evolution of the
stationary wave function $\psi^{\beta}(x,z)$.  \begin{itemize}

\item We can maintain the $\ttwo$ dependence of the functions $x_i$,
$a_i$ and $\alpha_i$ from the time dependent $\psi$.  This composition
of the KP flow with the bispectral involution is the bispectral flow
which was introduced in Section~\ref{sec:bisp-flow}.

\item Alternatively, since $\psi^{\beta}$ is a stationary wave
function of the KP hierarchy, one may add dependence on time so as to
make it a time dependent wave function.  Under this flow, the
functions $\lambda_i$, $a_i$ and $\alpha_i$ would become time
dependent while the $x_i$ would remain constant.  \end{itemize}

To avoid confusing these two flows of the wave function or the two
time dependencies of the functions $a_i$ and $\alpha_i$, the usual KP
flow of the function $\psi^{\beta}$ will be indicated by the variables
$T_n$ rather than the variables $t_n$.  That is, define the function
$\psi^{\beta}(x,\ttwo|\Ttwo,z)$ as follows: let the stationary wave
function $\psi(x,z)$ follow the second flow of the KP hierarchy until
time $\ttwo$, exchange the variables $x$ and $z$ in this new
stationary wave function, then let this function follow the second
flow of the KP hierarchy until time $\Ttwo$ (treating $\ttwo$ as a
constant).

The motion of the poles in $x$ of the function
$\psi^{\beta}(x,\ttwo|\Ttwo,z)$, under the variable $\Ttwo$ gives us
another Calogero-Moser particle system:
$(\vec{\BISxi}(\Ttwo),\vec{\BISeta}(\Ttwo))$.  In this way, we may
view $\beta$ as a map on the phase space of \cmpss\ given by
$\beta(\vec{x},\vec{y}) =(\vec{\BISxi},\vec{\BISeta})$.

Since the positions of the particles associated to a wave function are
given by the positions of the poles in $x$, it is clear that the
positions of the particles associated to $\psi^{\beta}$ are given by
the poles in $z$ of $\psi$. Therefore, we have $$\BISxi_i=\lambda_i.$$

By definition, we have
$\BISeta_i=\frac12\frac{d}{d\Ttwo}\lambda_i|_{\Ttwo=0}$.  However, the
next section will demonstrate that $\vec{\BISeta}$ is more easily
computed algebraically in terms of the dual grassmannian coordinates
$\lambda_i$ and $\gamma_i$.

\subsection{$\beta$ in terms of Dual Grassmannian
Coordinates}\label{sec:beta-dual}

As demonstrated above, the first component of
$(\vec{\BISxi},\vec{\BISeta})=\beta(\vec{x},\vec{y})$ is given by
$\vec{\BISxi}=\vec{\lambda}$ where $\lambda_i$ are the singular points
in the dual grassmannian coordinates of the solution associated to
$(\vec{x},\vec{y})$.  As a result of the two calculations described
below, we will similarly be able to determine $\vec{\BISeta}$ in terms
of the dual grassmannian coordinates.

Since the function $\psi(x,\ttwo,z)$ is a wave function for a solution
to the KP equation, it satisfies the non-stationary Schr\"odinger
equation $$ L\psi=\left(\frac{\partial}{\partial
\ttwo}-\frac{\partial^2}{\partial x^2}+\sum\frac{2}{(x-x_j)^2}\right)
\psi =0.  $$ Then for any $1\leq i\leq n$ the function
$(x-x_i)L\psi=0$, but its residue at the point $x=x_i$ is given by $$
\frac{a_i \dot
x_i}{q}+\frac{2za_i}{q}+2+\frac{2}{q}\sum_{j\not=i}\frac{a_j}{x_i-x_j}
$$ where $\dot x_i=\frac{d}{d\ttwo}x_i$.  Setting this equal to zero
and solving for $\dot x_i$, we find that $$ y_i=\frac12 \dot x_i = -
(z+\frac{\prod_{1\leq j\leq
n}(z-\lambda_j)}{a_i(z)}+\frac{1}{a_i(z)}\sum_{i\not=j}
\frac{a_j(z)}{x_i-x_j}).  $$

Then, considering $\psi^{\beta}$ as a function of $\Ttwo$, it is clear
that $$ \BISeta_i=-(x+\frac{\prod_{1\leq j \leq
n}(x-x_j)}{\alpha_i(x)}+\frac{1}{\alpha_i(x)}\sum_{i\not=j}
\frac{\alpha_j(x)}{\lambda_i-\lambda_j}).  $$

Alternatively, we can determine a similar formula for the dual
grassmannian coordinate $\gamma_i$.  Recall that the function
$\phi(x,z)=q(z)\psi(x,z)$ satisfies the conditions $c_i$ which
determine the solution.  In particular, $$
\phi'+\gamma_i\phi\bigg|_{z=\lambda_i}=0 $$ for all $x$ in the domain
of $\phi$, where differentiation is done with respect to the spectral
parameter $z$.  Consequently, $$
\frac{\phi'}{\phi}\bigg|_{z=\lambda_i}=-\gamma_i.  $$ Using the form
of $\psi$ which has been written in terms of the residues $\alpha_j$
in $z$, it is determined that $$ \gamma_i=-(x+\frac{\prod_{1\leq j
\leq
n}(x-x_j)}{\alpha_i(x)}+\frac{1}{\alpha_i(x)}\sum_{j\not=i}
\frac{\alpha_j(x)}{\lambda_i-\lambda_j}+\sum_{j\not=i}
\frac{1}{\lambda_i-\lambda_j}).  $$

Therefore, $$
\BISeta_i=\gamma_i+\sum_{j\not=i}\frac{1}{\lambda_i-\lambda_j}.  $$
This provides us with information about the image of $\beta$ as a map
from $\CM_0$.  In addition, it gives us the ability to determine
explicitly the dual grassmannian coordinates of a point under the
bispectral involution on a large class of bispectral rational
solutions.  These results are summarized by Theorem~\ref{image} and
Theorem~\ref{coords}.

\subsection{Linearization of Particle Systems}\label{sec:linear}

In this section, we will consider the bispectral flow of
$(\vec{\BISxi},\vec{\BISeta})$.  That is, given a particle system
$(\vec{x}(\ttwo),\vec{y}(\ttwo))$, we study
$(\vec{\BISxi}(\ttwo),\vec{\BISeta}(\ttwo))=
\beta(\vec{x}(\ttwo),\vec{y}(\ttwo))$.  Using
Theorem~\ref{image}, this flow can be determined simply by observing
the action of the KP flow on the parameters $\lambda_i$ and
$\gamma_i$.

The action of the KP flows on a condition of the form $$
f'(\lambda)+\gamma f(\lambda) $$ is determined simply in terms of
Definition~\ref{gamma+}.  Since the KP flows are isospectral, the
positions of the singular points, $\lambda_i$, are constant.  As shown
earlier, the positions of the particles, $\BISxi_i$, are given exactly
by the coordinates $\lambda_i$.  Consequently, the positions of the
particles, $\BISxi_i$, are fixed under the bispectral flow.
Furthermore, the result of the $n^{th}$ flow for time $t$ on the
parameter $\gamma$ is $\gamma\to n\lambda^{n-1} t +\gamma$.  As shown
earlier, we can determine $\vec{\BISeta}$ by the formula
$\BISeta_i=\gamma_i+\sum\frac{1}{\lambda_i-\lambda_j}$.  Since
$\gamma$ is a linear function of all time variables and the
$\lambda_i$ are constant, it is clear that the $\mu_i$ have been
linearized by $\beta$.  This proves Theorem~\ref{action-angle}.

The map $\sigma$ was originally considered only on the subset of
$\CM_0$ for which the parameters $x_i$ and $y_i$ are real numbers.
Recall\footnote{I have changed the notation slightly to agree with
that used in this paper.} that $\sigma(x_i,y_i)=(\Mxi_i,\Meta_i)$ was
defined so that $\Mxi_i=\lim_{\ttwo\to\infty}y_i$ is the asymptotic
velocity of the $i^{th}$ particle.  The map $\sigma$ can then be
related to $\beta$ using the fact that the eigenvalues of the matrix
$\Lambda$ are $\{-\lambda_i\}$ \cite{Kr} and, as always, are preserved
by the $\ttwo$ flow.  Then, since $$
\lim_{\ttwo\to\infty}\Lambda_{ij}=\lim_{\ttwo\to\infty}y_i\delta_{ij}
$$ we determine $-\Mxi=\lambda=\BISxi$.

Similarly, note that $\Meta$ is given as a linear function by the
formula $\Meta(\ttwo)=f'(\Mxi) \ttwo + \Meta(0)$ where the Hamiltonian
is given by $\hbox{tr}\,f(\Lambda)$ \cite{AMcM}~(Amplification~1).  In
this case $f(\Lambda)=\Lambda^2$, and so we get agreement with the
first coefficient of $\BISeta$ above.  That is, $$
-2\BISxi=-\frac{d}{d\ttwo}\BISeta=\frac{d}{d\ttwo}\Meta=2\Mxi.  $$
This is sufficient to prove Theorem~\ref{sigma}.  Thus, $\beta$
restricted to the domain of the map $\sigma$ is given simply by
$-\sigma+(0,\vec{c})$, where $\vec{c}$ is a constant of evolution.

\subsection{Examples}\label{sec:examps}

\subsubsection{Solving the System with the Bispectral Involution}

Consider the generic ``2 particle'' case given by $$
c_i(f)=f'(\lambda_i)+\gamma_if(\lambda_i)=0\qquad i=1,2 $$ and
$q(z)=(z-\lambda_1)(z-\lambda_2)$.  The parameters $\lambda_i$ and
$\gamma_i$ can form a matrix whose determinant is $\tau$ in two ways.
First, we can use the matrix $\M$ which gives $\tau$ in the gauge
given by $q=z^2$ and translate it to the bispectral gauge.  Then
\begin{eqnarray*} \tau(x,0,\ldots) &=& (x-x_1)(x-x_2)\\
 &=& e^{-(\lambda_1+\lambda_2)x} \det \M\\
 &=& e^{-(\lambda_1+\lambda_2)x} \det (c_i(z^j e^{xz}))\\
 &=& \det M_1 \end{eqnarray*} where $$
M_1=\left(\matrix{x+\gamma_1&x+\gamma_2\cr\lambda_1x+\lambda_1\gamma_1+1
& \lambda_2x+\lambda_2\gamma_2+1}\right) $$

On the other hand, we know that the eigenvalues of the Moser matrix
$\Lambda$ are simply $-\lambda_i$.  Since, the bispectral involution
exchanges $\lambda_i$ and $x_i$, we can also determine the
$\tau$-function as \begin{eqnarray*} \tau(x) &=& \det(xI+
\Lambda^{\beta})\\
 &=& \det
((x+\BISeta_i)\delta_{ij}+\frac{1-\delta_{ij}}{\BISxi_i-\BISxi_j})\\
 &=& \det
((x+\gamma_i+\sum\frac{1}{\lambda_i-\lambda_j})\delta_{ij}
+\frac{1-\delta_{ij}}{\lambda_i-\lambda_j})\\
 &=& \det M_2 \end{eqnarray*} where $$
M_2=\left(\matrix{x+\gamma_1+\frac{1}{\lambda_1-\lambda_2}&
\frac{1}{\lambda_1-\lambda_2}\cr\frac{1}{\lambda_2-\lambda_1}
&x+\gamma_2+\frac{1}{\lambda_2-\lambda_1}}\right).  $$
Note that the matrix $M_2$ is linear in time along the diagonal and
constant off of the diagonal.

The matrices $M_1$ and $M_2$, which arise in very different contexts
and are constructed in ways which do not appear similar, can be made
equal with only a few basic row operations.

\subsubsection{Determining the Action of the Bispectral Involution}

As an application of Theorem~\ref{coords}, the dual grassmannian
coordinates of the image under the bispectral involution of a specific
case of the previous example are determined below.

Let $\lambda_1=\gamma_1=1$ and $\lambda_2=\gamma_2=2$.  Then one may
determine that $$
\psi_1=\left(1+\frac{7+3x-3z-2xz}{(1+3x+x^2)(2-3z+z^2)}\right)e^{xz}
$$ $$ \left(\matrix{x_1\cr
x_2}\right)=\left(\matrix{\frac{-3+\sqrt{5}}{2}\cr
\frac{-3-\sqrt{5}}{2}}\right)
$$ and $$ \left(\matrix{y_1\cr
y_2}\right)=\left(\matrix{\frac{-6+\frac{1}{\sqrt{5}}+\sqrt{5}}{4}\cr
\frac{-6-\frac{1}{\sqrt{5}}-\sqrt{5}}{4}}\right).  $$

Then,the wave function of the solution given by the dual parameters
$\lambda_1=x_1$, $\lambda_2=x_2$, $\gamma_1=y_1-\frac{1}{x_1-x_2}$ and
$\gamma_2=y_2-\frac{1}{x_2-x_1}$ is $$
\psi_2=\left(1+\frac{7+3z-3x-2zx}{(1+3z+z^2)(2-3x+x^2)}\right)e^{xz}=
\beta(\psi_1).  $$


\noindent {bf Acknowledgements:} Thanks to my advisor, E. Previato,
for introducing me to the subject and the problems and providing me
with the knowledge to work on them.  Thanks to E.~Previato, D.~Fried
and D.~Blair for helpful discussions and ideas.  I am also grateful to
the referee for helpful comments and suggestions.

 \end{document}